\documentstyle[amstex,manuscript,osa]{revtex}

\begin{document}
\title{Ising-Bloch Transition in Degenerate\ Optical Parametric Oscillators}
\author{Isabel P\'{e}rez-Arjona, Fernando Silva, Germ\'{a}n J. de Valc\'{a}%
rcel, Eugenio Rold\'{a}n,}
\address{Departament d'\`{O}ptica, Universitat de Val\`{e}ncia,\\
Dr. Moliner 50, 46100 Burjassot, Spain.}
\author{V\'{\i}ctor J. S\'{a}nchez-Morcillo}
\address{Departament de F\'{\i}sica Aplicada, Universitat Polit\`{e}cnica de%
\\
Valencia,\\
Ctra. Nazaret-Oliva S/N, 46730 Grao de Gandia, Spain.}
\date{\today }
\maketitle

\begin{abstract}
Domain walls in type I degenerate optical parametric oscillators are
numerically investigated. Both steady Ising and moving Bloch walls are
found, bifurcating one into another through a nonequilibrium Ising--Bloch
transition. Bloch walls are found that connect either homogeneous or roll
planforms. Secondary bifurcations affecting Bloch wall movement are
characterized that lead to a transition from a steady drift state to a
temporal chaotic movement as the system is moved far from the primary,
Ising--Bloch bifurcation. Two kinds of routes to chaos are found, both
involving tori: a usual Ruelle-Takens and an intermittent scenarios.
\end{abstract}

\section{Introduction}

Nonlinear optical systems with broken phase symmetry and high Fresnel number
have a tendency to emit light beams whose transverse section displays
uniformly illuminated domains (usually spatially homogeneous) separated by
dark lines, the so-called domain walls (DWs). These structures are also
commonplace in self-oscillatory chemical reactors (like some variants of the
BZ reaction) forced at a 2:1 resonance, and in weakly damped nonlinear
mechanical systems (chains of coupled pendula, or fluids) when
parametrically forced. In both cases, sustained waves appear that oscillate
at half the driving frequency. An optical analogue of this is the degenerate
optical parametric oscillator (DOPO): a $\chi ^{(2)}$ cavity is driven by a
coherent light field of frequency $2\omega $ and the system starts to
oscillate (above a certain threshold) at the subharmonic frequency $\omega $.

Examples of DWs have been predicted to occur in the last few years in
several nonlinear optical resonators, such as DOPOs \cite%
{Longhi97,Trillo97,Staliunas98,LeBerre99,Rabbiosi02}, vectorial Kerr
cavities \cite{Izus00,Sanchez00b}, type II second harmonic generation \cite%
{Michaelis01} and, importantly, have also been experimentally realized in
parametric mixing \cite{Taranenko98,Larionova03}. Related phenomena have
also been reported in nascent optical bistability \cite{Tlidi98}, in which
case domains of low and high light intensity are separated (or better,
joined) by a switching front, and in single feedback mirror experiments in
the presence of an intrinsic polarization instability \cite{Grosse00}, where
domain patterns are observed.

Usually a DW asymptotically joins two homogeneous states, $u_{\pm }$. In the
simplest, most usual case, $u_{\pm }$ have the same amplitude and opposite
phase, i.e. $u_{+}=-u_{-}$, corresponding to two antisymmetric fixed points
in the complex plane $\left\langle 
\mathop{\rm Re}%
u,%
\mathop{\rm Im}%
u\right\rangle .$ In this representation, a trajectory connecting the two
domains can follow two paths, corresponding to two different types of walls:
either crossing the complex zero (Ising wall) or surrounding it (Bloch
wall). In terms of the phase, in an Ising wall there is a discontinuous
variation of the field phase across the wall whereas in a Bloch wall the
phase angle rotates through $\pi $ across the wall. This is a crucial
difference between Ising and Bloch walls as it implies that the latter are
chiral while the former are not. The chirality measures the direction and
magnitude of the rotation of the phase angle at the wall core and, as two
directions of rotation of the phase angle can in principle be expected,
Bloch walls of positive and negative chirality can exist. Alternatively, in
an Ising wall the field intensity vanishes at its core, whilst it is
minimum, but not zero, in a Bloch wall. That is the reason why in nonlinear
optics Ising and Bloch walls are sometimes referred to as dark and grey
solitons, respectively.

In nonequilibrium systems, Ising walls bifurcate into Bloch walls by varying
a parameter of the system, the bifurcation point corresponding to the
nonequilibrium Ising-Bloch transition \cite{Bulaevskii63} (NIB in the
following). In systems showing this behaviour, the chirality behaves as an
order parameter, and can be described, in principle, by a nonlinear
evolution equation: the NIB transition is then related to a bifurcation of
the chirality parameter \cite{Coullet90}. Importantly, in gradient systems
(those whose dynamics derives from a potential), both Ising and Bloch walls
are at rest. This is a consequence of the equivalence between the states
connected by the wall, characterized by the same free energy. However, in
nongradient systems, such as the DOPO, a generic property of Bloch walls is
that they move with a velocity proportional to (or related with) their
chirality, while Ising walls are at rest \cite{Coullet90}.

The NIB transition has been found in systems of very different nature, such
as nematic liquid crystals \cite{Frisch94}, and reaction-diffusion systems 
\cite{Hagberg93}. In the context of nonlinear optics, it has been found in
type II optical parametric oscillators when cavity birefringence and/or
mirror dichroism are taken into account \cite{Izus00}, and in type II second
harmonic generation \cite{Michaelis01}. It has also been studied in
universal equations describing nonlinear optical cavities in some limiting
cases, as it is the case of the parametrically driven Ginzburg--Landau
equation \cite{Skryabin} and the parametrically driven nonlinear Schr\"{o}%
dinger (PDNLS) equation \cite{deValcarcel}. In \cite{Michaelis01} a
universal criterion for evaluating the NIB transition boundary in a wide
variety of nongradient systems was proposed.

The results obtained in \cite{deValcarcel} are particularly relevant for the
present work. Whilst it is well known that the PDNLS equation\ with
self--defocusing nonlinearity exhibits domain walls ($\tanh $ solitons) \cite%
{Elphick}, it was not known until recently \cite{deValcarcel} that domain
walls are also solutions of this equation when the nonlinearity is
self--focusing (in which case the basic solitonic structure is a $%
\mathop{\rm sech}%
$). This fact makes possible that the same nonlinear system exhibits both
bright and dark solitons in adjacent regions of the parameter space, as
bright solitons are stable solutions of the PDNLS equation \cite{Miles}.
Moreover, in \cite{deValcarcel} it was shown that a NIB transition occurs in
this equation. It is to be remarked that the DWs found in the self--focusing
PDNLS equation can connect not only homogeneous states, as usual domain
walls do, but also patterned states. As that equation has been derived in
nonlinear optics for a number of systems (in particular for DOPOs with large
pump detuning \cite{Longhi97}, vectorial Kerr cavities with large cavity
anisotropy \cite{Sanchez00b}, and fiber rings with phase-sensitive
amplification \cite{Mecozzi}) the results presented in \cite{deValcarcel}
imply that these systems should exhibit domain walls and a NIB transition in
some parameter range.

In this paper we show numerically that the Ising walls previously studied in
the type I DOPO can experience the NIB transition for positive signal
detuning. Remarkably, we show that this occurs in a parameter domain for
which the PDNLS equation, which was derived for large pump detuning \cite%
{Longhi97}, cannot be applied. The moving Bloch fronts that appear beyond
the NIB bifurcation connect two homogeneous states with opposite phase and
exist in a finite parameter region. By varying a control parameter (e.g.,
the signal detuning) the homogeneous solutions connected by the Bloch wall
become modulationally unstable and are replaced by rolls. These new domain
walls exhibit a very rich dynamic behaviour that we analyze in detail. We
would like to mention that some of the results presented here have been
previously found by Le Berre {\it et al.} \cite{LeBerre99} in a porpagation
model for DOPO. In particular, they pointed out that DWs exist in the DOPO
also for positive signal detuning, that these domain walls can connect
patterned states, and it was also speculated by the authors that the system
could exhibit a NIB transition. The main goals of the present work are to
characterize the NIB transition in DOPOs and to describe in detail the
nonlinear dynamics of the Bloch walls. Our study, being specific of the DOPO
model, finds however a qualitative parallelism with our previous work on the
PDNLS equation \cite{deValcarcel}; we thus conjecture that our findings
should be applicable to other systems.

\section{Model, homogeneous solutions and their stability}

The standard, mean-field model for a type I DOPO reads \cite{Oppo94} 
\begin{eqnarray}
\frac{\partial A_{0}}{\partial T} &=&\gamma _{0}\left[ -(1+i\Delta
_{0})\left( A_{0}-E\right) -A_{1}^{2}+ia_{0}\nabla ^{2}A_{0}\right] ,
\label{m1} \\
\frac{\partial A_{1}}{\partial T} &=&\gamma _{1}\left[ -(1+i\Delta
_{1})A_{1}+A_{0}A_{1}^{\ast }+ia_{1}\nabla ^{2}A_{1}\right] ,  \label{m2}
\end{eqnarray}
where $A_{n}$ are the slowly varying envelopes of the intracavity pump $%
(n=0) $ and signal $(n=1)$ fields, $\gamma _{n},\Delta _{n}$ and $a_{n}$
their corresponding cavity decay rates, detuning and diffraction
coefficients, and $E$ is the amplitude of the injected pump, which in
general may depend on transverse coordinates. The phase-matching condition
imposes that $\gamma _{0}a_{0}=2\gamma _{1}a_{1}$. In this paper we shall
consider the 1D limit of Eqs. (1). This situation corresponds, e.g., to a
slab waveguide configuration for the resonator, in which the fields are
confined in one direction of the transverse plane, say $Y$, the diffraction
acting only in the $X$ direction, and then $\nabla ^{2}=\partial
^{2}/\partial X^{2}$. In the following we shall concentrate in the
particular case $\gamma _{0}=\gamma _{1}\equiv \gamma $ (hence $a_{1}=2a_{0}$%
), and use normalized time $t=\gamma T$ and space $x=X/\sqrt{a_{1}}$.
Moreover, throughout this paper we shall assume that the two fields
oscillate within the same cavity and then we shall keep the relation $\Delta
_{0}=2\Delta _{1}$ fixed.

For a spatially uniform injected pump, Eqs.(\ref{m1}) and (\ref{m2}) admit
two homogeneous solutions. The trivial solution $\left\{
A_{0}=E,A_{1}=0\right\} $ exists for all parameter sets, while the
''lasing'' solution 
\begin{gather}
\left| A_{1}\right| ^{2}=\Delta _{0}\Delta _{1}-1\pm \sqrt{E^{2}(1+\Delta
_{0}^{2})-(\Delta _{0}+\Delta _{1})^{2}},\,  \nonumber \\
\left| A_{0}\right| ^{2}=1+\Delta _{1}^{2},  \label{homog}
\end{gather}
that corresponds to the subharmonic (signal) generation, requires a
threshold pump. The bifurcation from the trivial solution to the lasing
solution (\ref{homog}) occurs at $E=\sqrt{1+\Delta _{1}^{2}}$; it is
subcritical (and then the lasing solution can coexist with the stable
trivial state) when $\Delta _{0}\Delta _{1}>1$, and is supercritical in the
opposite case \cite{Lugiato88}. For positive $\Delta _{1}$ (which is the
case we consider throughout the paper) this is the only bifurcation that
affects the trivial state. Note that the discrete phase symmetry $%
A_{1}\rightarrow -A_{1}$ supported by Eqs.(\ref{m1}) and (\ref{m2}) makes
that two equivalent solutions of equal intensity but opposite phase exist
and are dynamically equivalent, and this opens the possibility of exciting
DWs connecting them. For negative $\Delta _{1}$ (a case we do not discuss
here) a pattern forming instability affecting the trivial state occurs at $%
E=1$ \cite{Oppo94} leading to the appearance of roll patterns \cite%
{deValcarcel96}.

The stability of the lasing solution (\ref{homog}) against space-dependent
perturbations has been also investigated \cite{Longhi97,Staliunas98}. A
linear stability analysis predicts that a pattern forming instability of the
homogeneous state\ occurs for pump values below a critical value or,
alternatively, above a critical detuning value. The analytical expressions
are rather involved, but the instability threshold can be found in
particular cases numerically. In Fig. 1 we represent the domain of existence
of the different solutions in the plane $\left\langle \Delta
_{1},E\right\rangle $. The homogeneous solution exists above line (a), being
stable to the left of the dashed line (c) and unstable versus the roll
pattern to the right of line (c). The roll pattern appearing in this region
has a dc component, at difference with the roll pattern appearing for
negative detuning, which has no dc component and thus its visibility equals
unity. In the region marked as BS, between lines (a) and (b), the trivial
and homogeneous solutions coexist but as the latter is modulationally
unstable, in this region bright cavity solitons are formed \cite%
{Longhi97,Staliunas97}.

\section{Nonequilibrium Ising--Bloch transition}

In this section we report the results of our numerical study of the DW
dynamics. We integrated numerically Eqs. (\ref{m1}) and (\ref{m2}) making
use of a split-step algorithm with periodic boundary conditions. Spatial
grids from $1024$ to $8192$ points were used in order to check convergence.
Results reported here were obtained with an integrating window length $L=316$
(other values were also used) and the temporal step was lowered up to $%
\delta t=10^{-2}$ in order to yield $\delta t-$independent results. We
considered two cases for the pump amplitude profile $E$ corresponding to (i)
an infinitely extended plane-wave field and (ii) a flat but spatially
limited field, flat in order to avoid any gradient effect on the DW
dynamics. For this last case we chose 
\begin{equation}
E=E_{0}\exp \left[ -\left( x/\Delta x\right) ^{8}\right] ,  \label{E(x)}
\end{equation}%
with $\Delta x=0.45L$, which is top-hat like: it is flat around $x=0$ and
null close to the border of the integration region $\left( x=\pm L/2\right) $%
. This supergaussian profile is used with the only purpose of simulating a
flat and finite pump and that is not essential for the results here
reported. In fact very similar results, differing mainly on minor
quantitative details, are found in the two studied cases, thus demonstrating
the robustness of the reported behaviour.

We first report the results obtained for a top-hat pump. A wall connecting
the two homogeneous, oppositely phased states was excited by using an
appropriate initial condition. The two kinds of walls, Ising and Bloch, were
found at different parameter values. Whilst the intensity trace $\left|
A_{1}\left( x\right) \right| ^{2}$ of the walls formed in the subharmonic
field does not allow to distinguish clearly between Ising and Bloch walls
-due to the very small value of the intensity at the core of the wall- the
different nature of the walls can however be clearly appreciated in a
parametric representation $\left\langle 
\mathop{\rm Re}%
A_{1},%
\mathop{\rm Im}%
A_{1}\right\rangle $ of these interfaces. In Fig. 2 the parametric
representations of both an Ising wall, Fig. 2(a), obtained for $E_{0}=3$ and 
$\Delta _{1}=1.2$, and a Bloch wall, Fig. 2(b), obtained for $E_{0}=3$ and $%
\Delta _{1}=1.5$, are shown. The NIB transition, numerically computed from
Eqs.(\ref{m1}) and (\ref{m2}), is plotted in Fig. 1, line (d).

The NIB\ transition can be clearly identified as a bifurcation of a
chirality parameter defined as \cite{deValcarcel} 
\begin{equation}
\chi =%
\mathop{\rm Im}%
\left[ A_{1}^{\ast }\left( x_{0}\right) \partial _{x}A_{1}\left(
x_{0}\right) \right] ,  \label{chirality}
\end{equation}
where $x_{0}$ is the position of the wall core, i.e{\it .} where $\left|
A_{1}\right| $ is closest to zero. Notice that if we use a polar
decomposition of $A_{1}\left( x\right) $ as $A_{1}\left( x\right) =\left|
A_{1}\left( x\right) \right| \exp \left[ i\phi _{1}\left( x\right) \right] $%
, then $\chi =\left| A_{1}\left( x_{0}\right) \right| ^{2}\partial _{x}\phi
_{1}\left( x_{0}\right) $. Thus this quantity is sensitive to both the sense
and magnitude of the rotation of the phase angle at the wall, $\partial
_{x}\phi _{1}\left( x_{0}\right) $, and to the intensity at the wall core, $%
\left| A_{1}\left( x_{0}\right) \right| ^{2}$. Consequently $\chi =0$ for
Ising walls and $\chi \neq 0$ for Bloch walls, and the NIB can be understood
as a pitchfork bifurcation of the chirality parameter.

As Eqs. (\ref{m1}) and (\ref{m2}) are invariant under spatial translations $%
\left( x\rightarrow x+x^{\prime }\right) $ and spatial reflections $\left(
x\rightarrow -x\right) $, if a Bloch wall $A_{1}^{\left( 1\right)
}=A_{B}\left( x-x_{0},t\right) $ exists with chirality $\chi _{1}$ another
Bloch wall $A_{1}^{\left( 2\right) }=A_{B}\left( x_{0}-x,t\right) $ also
exists and has chirality $\chi _{2}=-\chi _{1}$. This is important, as a
generic property of domain walls in nongradient systems (at least for domain
walls of the $\tanh $ type) is that they move with a velocity which is
proportional to their chirality \cite{Coullet90}, so that Bloch walls with
opposite chiralities move in opposite directions, and Ising walls $\left(
\chi =0\right) $ are at rest.

In Fig. 3 the signal field intensity $\left| A_{1}\left( x\right) \right|
^{2}$ is shown for the same set of parameters as in Fig. 2(a), and the Ising
wall is clearly appreciated in the center of the illuminated region, where
it remains at rest indefinitely. In Fig. 4(a) a similar representation for
the same parameters as in Fig.2 (b), that is, for a Bloch wall, is shown. In
this case the position of the wall varies with time as can be seen in Fig.
4(b), which displays the time evolution of the wall position: contrarily to
the Ising wall, the Bloch wall drifts at a velocity $v$ (that depends on the
parameter values) until it reaches the boundary of the pumped area, where it
remains locked.

Let us notice that, in the wall neighbourhood, the intensity profile of the
signal field shows spatial oscillations that decay exponentially to the
homogeneous state given by (\ref{homog}), see Figs. 3 and 4. These spatial
modulations appear in the parametric representation as a spiralling of the
heteroclinic trajectory around the stable fixed points representing the
homogeneous stable solutions, see Fig. 2. These modulations have a
characteristic wavenumber that can be predicted by a spatial stability
analysis \cite{Sanchez00}, and are typical of diffraction-supported fronts.
These modulations play an important role in the stabilization of the walls,
and allow for the existence of bounded states \cite{Sanchez00}. The number
of orbits in the parametric representation increases when approaching the
pattern forming instability boundary, line (c) in Fig. 1.

\section{Nonlinear dynamics of the Bloch wall}

We consider next the dynamical properties of Bloch walls. First we consider
the case of the top-hat pump profile of Eq.(\ref{E(x)}) and then we consider
the case of plane--wave pumping. In both cases, a typical route is reported,
as obtained for $E=3$, and the signal detuning $\Delta _{1}$ is left as the
control parameter.

\subsection{Pump field with finite transverse extension}

The Ising wall represented in Fig. 3 (also in Fig. 2 (a)), corresponding to $%
\Delta _{1}=1.2$, suffers instabilities at its core by increasing $\Delta
_{1}$. For the selected pump value, the NIB transition is found at $\Delta
_{1}=1.362$. At this point the value of the intensity at the dip becomes
different from zero, although it has an extremely small value, and the wall
starts to move with a constant drift, as shown in Fig. 4 for $\Delta
_{1}=1.5 $. We note that the drift velocity of the wall is not exactly
constant as it slightly oscillates with a very low frequency. We shall
return to this point in the next section.

For detuning values larger than $\Delta _{1}=1.73$ the homogeneous solution (%
\ref{homog}) becomes modulationally unstable, see Fig. 1, and each of the
two domains connected by the wall develop extended spatial oscillations.
Bloch walls still exist in this parameter region but now they connect two
patterned states instead of two homogeneous solutions \cite{LeBerre99}. This
type of domain wall was already found in the PDNLS \cite{deValcarcel}. An
example of the intensity profile of a wall affected by the modulational
instability is shown in Fig. 5, corresponding to $\Delta _{1}=2.01.$ For
this value of the detuning, after a transient the wall moves at nearly
constant velocity until it reaches the border of the illuminated region,
where it remains locked, see Fig. 5(b), similarly to the case of Fig. 4(b).
Notice however that now the wall is bounced from the border of the
illuminated region before it locks to it. Further increasing the detuning,
the Bloch wall keeps existing but exhibits an erratic motion until finally
it is also locked to the boundary. This ''chaotic'' motion of the Bloch
walls is observed for $\Delta _{1}>2.03$.

We see that with a pump of finite extension, the dynamics of the domain
walls is always transient as Bloch walls get eventually locked to the
boundary of the illuminated region. Then in order to quantitatively analyze
the temporal dynamics of Bloch walls, it is necessary to consider an
infinitely extended plane--wave pump in order to avoid the influence of
boundary effects. This is done in the next subsection.

\subsection{Infinitely extended pump field}

For periodic boundary conditions and uniform pump, an even number of DWs had
to be excited. In our simulations, two DWs were initially created.
Consequently care had to be taken that the two DWs did not interact as the
dynamics of a single DW was to be studied. This was done by checking that
the two domain walls were far apart each other and that they did not
approach significantly during the numerical run. We analyzed the temporal
series of both the chirality and velocity of DWs. In Fig. 6 the dependence
of both quantities on the detuning $\Delta _{1}$ are shown for a pump value $%
E=3$. Next we analyze this bifurcation diagram in detail.

The NIB occurs at $\Delta _{1}=1.362$, as in the case of spatially limited
pump, and at this point both the chirality and velocity of the wall become
non--null through a pitchfork bifurcation. Nevertheless, for detunings
slightly below this value there appears some subtle dynamics, as
aniticipated. For a detuning of $\Delta =1.361$ we have observed that the
position of the DW oscillates in time, with null mean displacement and a
very small amplitude in the oscillations (of the order of a fraction of a
pixel - the determination of the intensity minimum was done through a
three-point parabolic fit). In Fig.7 (a) the power spectrum of the chirality
time series corresponding to this case is shown. The frequency of the
oscillation is so small ($f=1.793\cdot 10^{-4}$) that runs with a duration $%
\Delta t=2^{18}$ (more than 26 milion time steps) had to be considered in
order to have $47$ oscillations. Given the tiny value of the fundamental
frequency, power spectra were obtained after sampling every time unit. In
Fig. 7(b) the power spectrum corresponding to a detuning $\Delta _{1}=1.365$%
, that is beyond the NIB transition, is shown. The fundamental frequency in
the spectrum is larger than that in Fig. 7(a) and, remarkably, the dynamics
is richer as higher order harmonics appear. In the range of detunings $%
1.362<\Delta _{1}<1.85$, the movement of the wall is practically steady but
for these very low frequency, smallest amplitude, oscillations; that is, for
constant detuning the velocity is very nearly constant, its value increasing
with the detuning as can be seen in Fig. 6. Unfortunately, as the NIB
transition is crossed, it is no more possible to track the low frequency
dynamics as new oscillatory bifurcations involving much higher frequencies
develop. Thus, up to $\Delta _{1}=1.85$ the velocity of the Bloch wall is
not constant but the frequency of the oscillation, as well as its amplitude,
is so small that one can consider it as practically constant. In the
following we neglect this low frequency dynamics.

At $\Delta _{1}=1.85$ there is a clear Hopf bifurcation at which the
velocity is no more constant but begins to oscillate in time (in Fig. 6 we
plot both the maximum and minimum values of the velocity and chirality in
this oscillatory regime). In Fig. 8(a) the spectrum corresponding to that
clear periodic motion is shown. Let us remark that the frequency of these
oscillations is larger, by a factor greater than $350$, than the low
frequency oscillations above commented.

It is most interesting to notice that this Hopf bifurcation occurs at
exactly the same detuning value at which the homogeneous solution of the
system becomes modulationally unstable, line (c) in Fig.1. This does not
seem to be accidental or specific of this system as we have observed the
same phenomenon in the PDNLS equation \cite{deValcarcel,Perez}. Then it
seems that the appearance of a Hopf bifurcation in the movement of the Bloch
wall is forced by the modulational instability that the homogeneous solution
undergoes. Up to some extent this fact can be understood intuitively. The
appearance of a pattern on the background where the domain wall exists
necessarily influences the movement of the domain wall as the displacement
of the wall forces local changes in the spatial frequency of the pattern. On
their turn, these local readjustments in the pattern shape modify the
velocity of the Bloch wall. This is consistent with the fact that the
oscillations developed by the wall velocity increase their amplitude as the
detuning is increased, as it occurs with the modulation of the roll pattern
on which the wall is ''written''.

Further increasing the detuning leads to the appearance of new frequencies
in the spectrum. At $\Delta _{1}=1.93$, a new and incommensurate frequency
appears, see Fig. 8(b), reflecting the fact that the movement of the wall is
now quasiperiodic. At $\Delta _{1}=1.95$, there still appears a third new
incommensurate frequency, and then the movement of the wall corresponds to a
3--torus dynamics, Fig. 8(c).This quasiperiodic dynamics remains up to $%
\Delta _{1}=2.014$. In Fig. 9 we represent a projection of the attractor on
the $\left\langle v,\chi \right\rangle $ plane for a periodic and a
quasiperiodic behaviour, see caption. Notice in Fig. 6(a) that at $\Delta
_{1}\approx 1.96$, the minimum value of the chirality becomes negative,
although the velocity remains positive along all the depicted lines (see
also Fig. 9(b)). This means that, under oscillatory dynamics, there is no
more a direct relation between the sign and magnitude of the chirality and
those of the velocity.

We find it important to emphasize that the observed dynamics is due to the
existence of a DW: we have checked that the roll pattern existing for $%
\Delta _{1}>1.73$ is always steady for $E=3$. Then, the secondary
bifurcations affecting the Bloch wall movement are not induced by any
temporal dynamics of the pattern, but have to be attributed to the interplay
between the change in the wall velocity with detuning, on the one side, and
the changes in the spatial frequency and modulation of the roll pattern, on
the other side.

For detuning values within the range $2.014<\Delta _{1}<2.017$ (within the
shaded area depicted in Fig. 6) the dynamic behaviour of the wall cannot be
tracked as the two excited walls always coalesce into a single structure (a
cavity soliton). Although we tried to avoid this fact by changing the
initial conditions (more specifically, the initial chirality value of the
two walls) we have not been able of isolating the dynamics of a single
structure in this smallest domain of detunings.

Further, at $\Delta _{1}=2.017$, there appears a new qualitative change in
the dynamics of the Bloch wall: the time evolution of the chirality (or the
velocity) is again periodic but now the dynamics correspond to an attractor
different from the one shown in Fig. 9, as can be seen in Fig. 10(a). This
new attractor is two sided, that is, the orbit on the $\left\langle v,\chi
\right\rangle $ plane surrounds alternatively the two unstable symmetric
fixed points differing in the sign of the chirality and the velocity. As in
the previous case, this periodic attractor transforms into a chaotic one,
Fig. 10(b), after a series of bifurcations involving tori, see Fig. 11.
However, in this case the route to chaos is quite unusual: By increasing the
detuning, this motion remains periodic until $\Delta _{1}=2.0250$, then at $%
\Delta _{1}=2.0251$ the motion becomes quasiperiodic and at $\Delta
_{1}=2.0252$ an intermittent route to chaos (built on the torus) is
initiated. Close to this border laminar phases are very long and become
shorter as we move from the bifurcation, as usual. An example of this
intermittent behaviour is shown in Fig. 12 where a series of chirality
extrema is depicted. We have not tried to characterize this highly
complicated behaviour and just note that the same type of intermittencies
were found by some of us in laser models\cite{deValcarcel95,Redondo}.
Further increasing detuning laminar phases become progressively shorter
until eventually they disapear (at some detuning value between $2.035$ and $%
2.039$) the wall motion becoming chaotic. For still larger values of the
detuning, the dynamics of the wall remains chaotic until line (b) in Fig.1
is crossed as at this point the pattern is no more stable and bright ($%
\mathop{\rm sech}%
$ type) solitons appear after a transient.

\section{Conclusions}

We have presented numerical evidence of the existence of a nonequilibrium
Ising-Bloch transition in a type-I DOPO model with one transverse dimension.
Bloch walls have been found in two different forms, either connecting
homogeneous states, which are the usual ones, or connecting modulated
states, a kind of domain wall already found in the propagation model for
DOPO \cite{LeBerre99}, and in the parametrically driven, damped nonlinear
Schr\"{o}dinger equation \cite{deValcarcel}.

Bloch walls move and their movement undergoes complicated secondary
bifurcations eventually leading to chaotic motion involving
quasiperiodicity. Two routes to chaos have been described: A usual
quasiperiodic (Ruelle-Takens) scenario and an unusual quasiperiodic
intermittency. We have seen that the nonlinear dynamics of the wall movement
is related to the appearance of a modulational instability on the pattern on
which the domain wall is written. Let us remark that although the NIB
transition in DOPO has been reported for the particular case $\Delta
_{0}=2\Delta _{1}$, we have observed similar phenomena for different sets of
detuning values. In fact, for the case of large pump detuning (and signal
detuning of order one), the DOPO\ equations can be approximated by the PDNLS
equation \cite{Longhi97}, where the NIB\ transition as well as the nonlinear
dynamics of the Bloch wall have been previously reported by some of us \cite%
{deValcarcel,Perez}.

We would like to finish with a brief mention to the case of two transverse
spatial dimensions. In 2D the situation is by far much more complicated than
the 1D case we have analyzed here, as in 2D the dynamics of DWs is affected
not only by the NIB but also by curvature effects \cite{Izus00,LeBerre99}.
Moreover, boundary effects become determinant in two dimensions as domain
walls tend to be perpendicular to the boundary of the illuminated region.
Then the analysis of the NIB in a two dimensional optical system still
remains to be characterized.

We thank Javier Redondo (Departament de F\'{\i}sica Aplicada, Universitat
Polit\`{e}cnica de Val\`{e}ncia) for useful discussions about the processing
of the numerical data. This work has been supported by the Spanish
Ministerio de Ciencia y Tecnologia and the European fonds FEDER under
projects BFM2002-04369-C04-01 and BFM2002-04369-C04-04.

{\Large Figure captions}

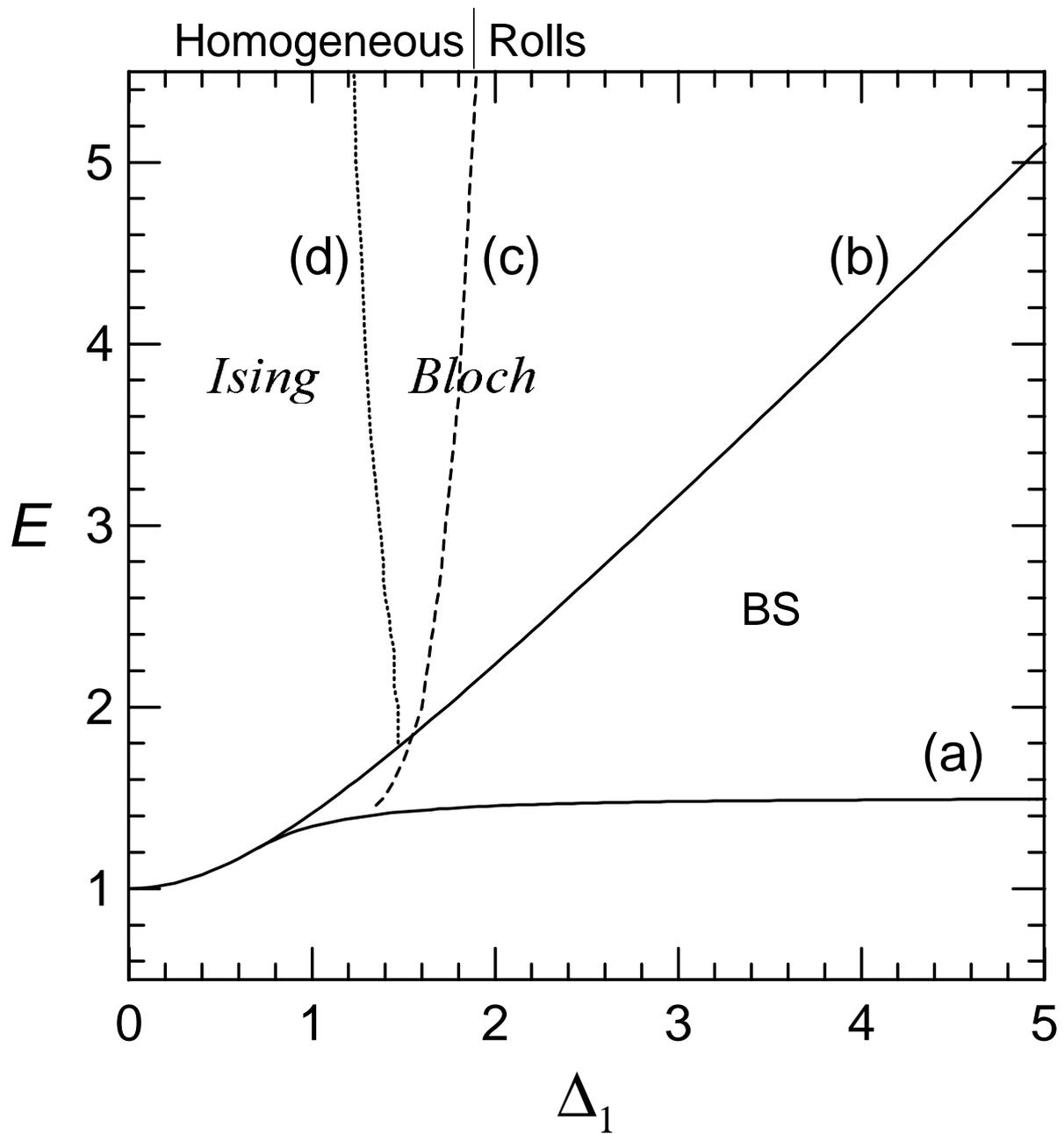
Fig. 1. Bifurcation diagram in the parameter space pump amplitude ($E$) vs.
signal detuning ($\Delta _{1}$).\thinspace The different lines denote: (a)
the location of the turning point of the homogeneous, ''lasing'' solution;
in (b) the trivial solution losses its stability; in (c) the homogeneous
solution (that is stable to the left of this line and unstable to the right)
suffers a modulational instability that gives rise to the appearance of
patterns; finally, line (d)\ denotes the location of the Ising--Bloch
transition. The region denoted by BS corresponds to the bistability domain
where bright solitons can be excited.

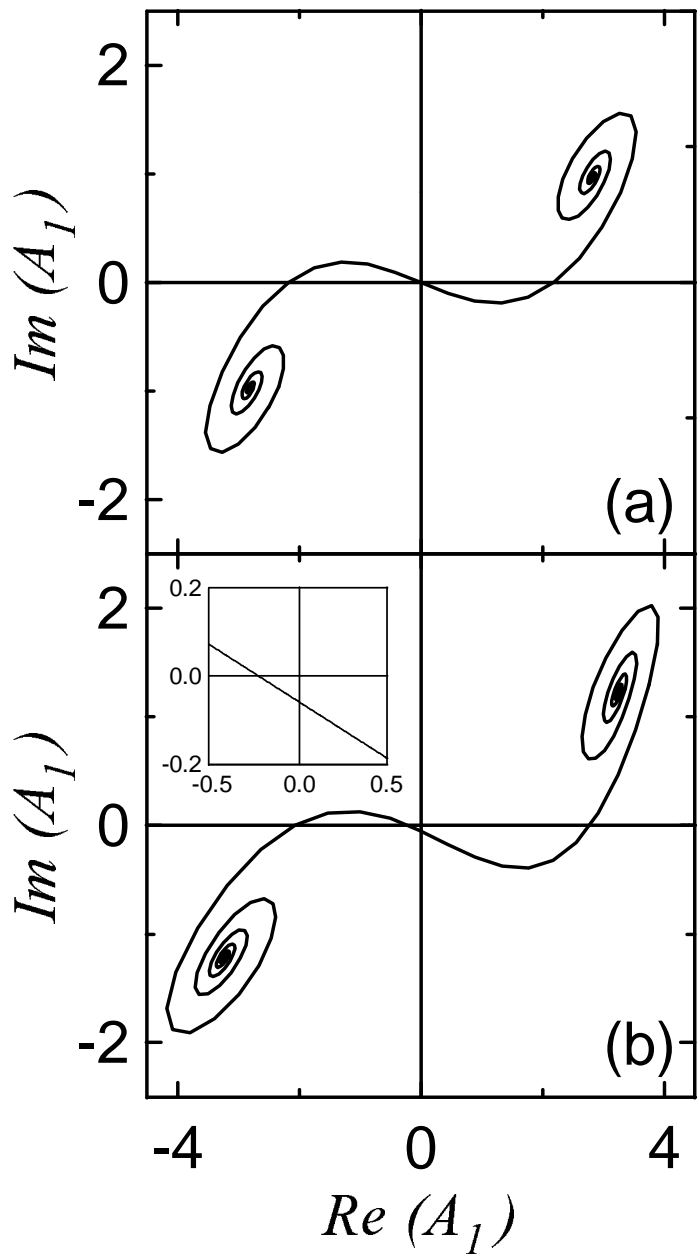
Fig. 2. Parametric representation of an Ising (a) and a Bloch (b) wall,
computed numerically from Eqs. (1) with $E=3$ and (a) $\Delta _{1}=1.2$, and
(b) $\Delta _{1}=1.5$.

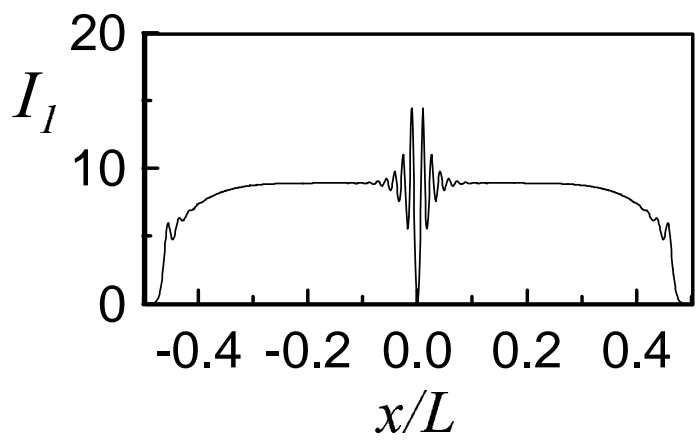
Fig. 3. Intensity profile of the signal field, $\left\vert A_{1}\right\vert
^{2}$, as a function of the normalized position $x/L$ for $E=3$ and $\Delta
_{1}=1.2$. An Ising wall appears at the center.

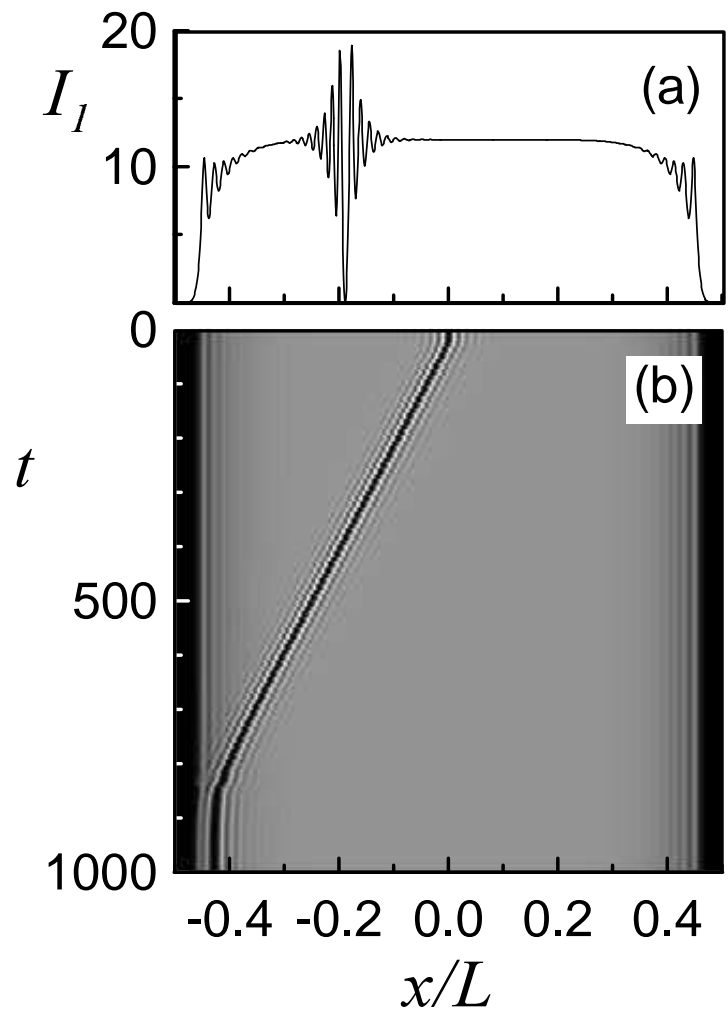
Fig. 4. (a) As in Fig.3 but for $E=3$ and $\Delta _{1}=1.5$ showing a Bloch
wall. (b) Time evolution of the pattern. The pattern in (a) corresponds to $%
t=400$ in the time evolution shown in (b).

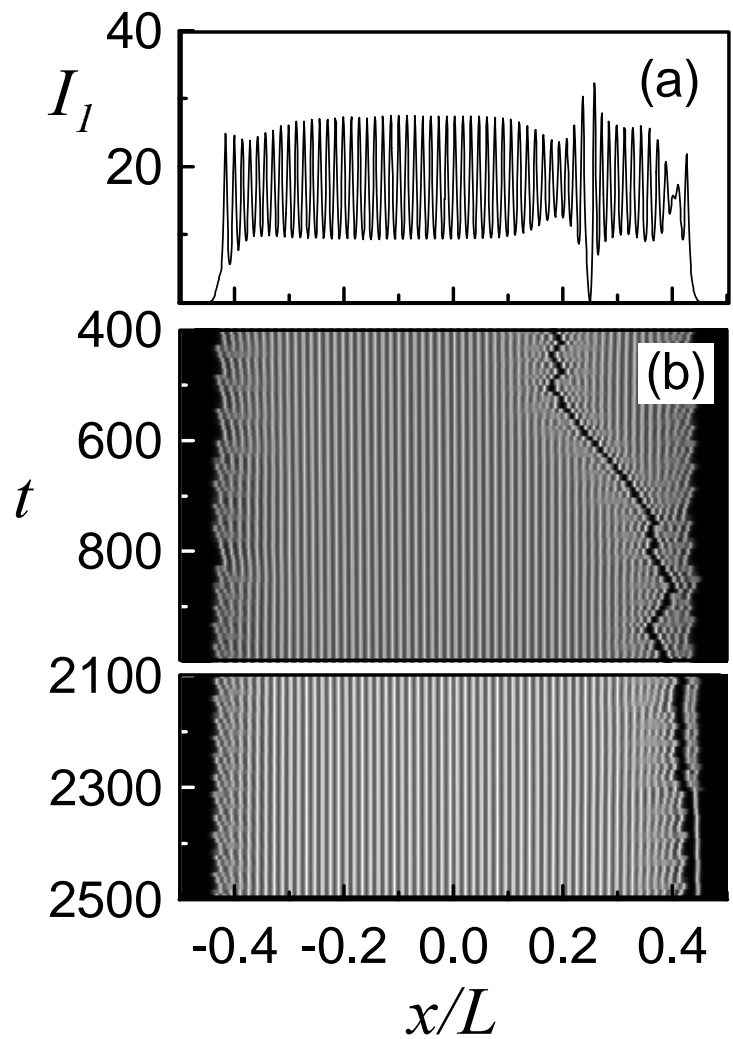
Fig. 5. As in Fig.4 but for $E=3$ and $\Delta _{1}=2$.

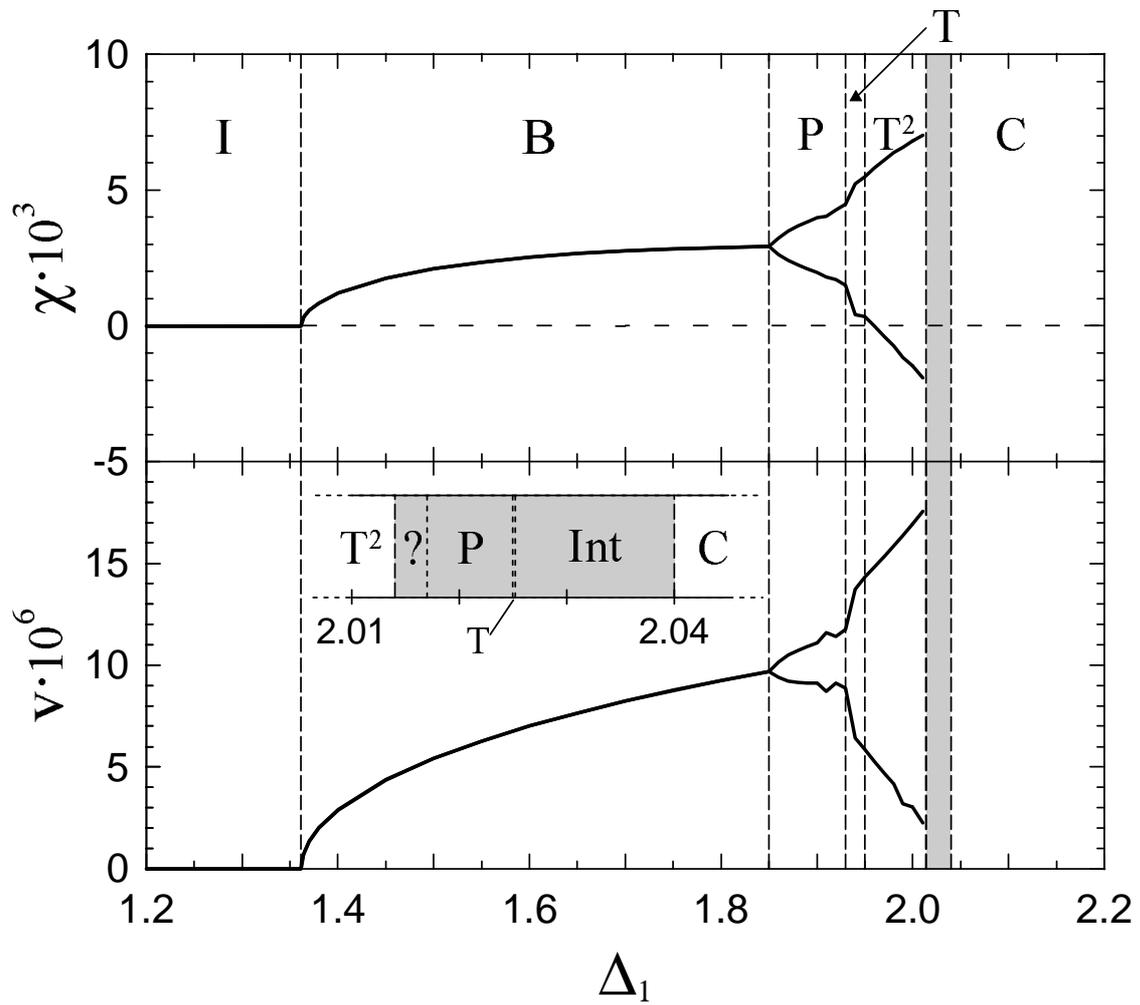
Fig. 6. Bifurcation diagram for the Chirality (a) and velocity (b) of the
wall as a function of detuning for $E=3$. $I$ denotes the region of
existence of Ising walls and $B$ the region of existence of Bloch walls with
constant velocity. In $P$ the motion of Bloch walls is periodic
(quasiperiodic in $T$ and $T^{2}$, and chaotic in $C$) and the maximum and
minimum value of the chirality and velocity are represented. The inset
diagram show the dynamics in the grey region, where $Int$ correspond with
the intermittent route to chaos and the symbol ($?$) marks the detuning
where the dynamics of a single wall cannot be tracked (see text).

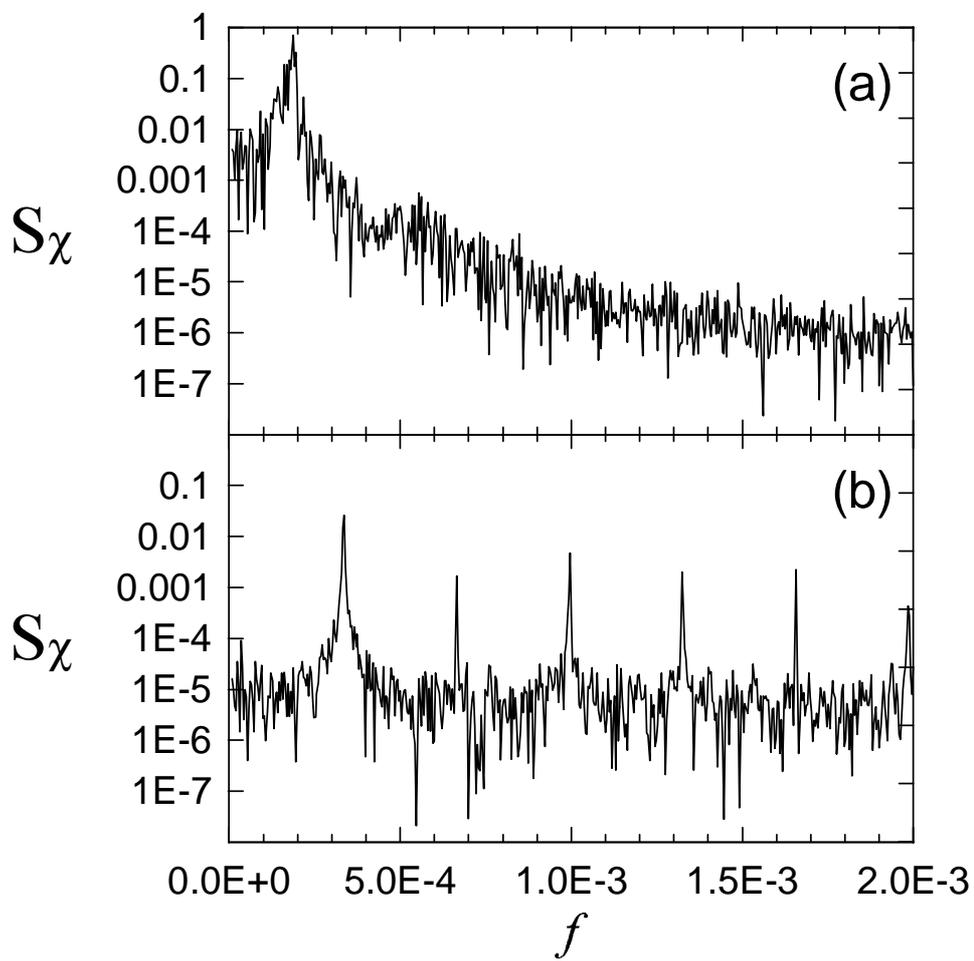
Fig. 7. Chirality spectrum for the time series corresponding to $E=3$, $%
\Delta _{1}=1.361$ (a) and $\Delta _{1}=1.365$ (b).

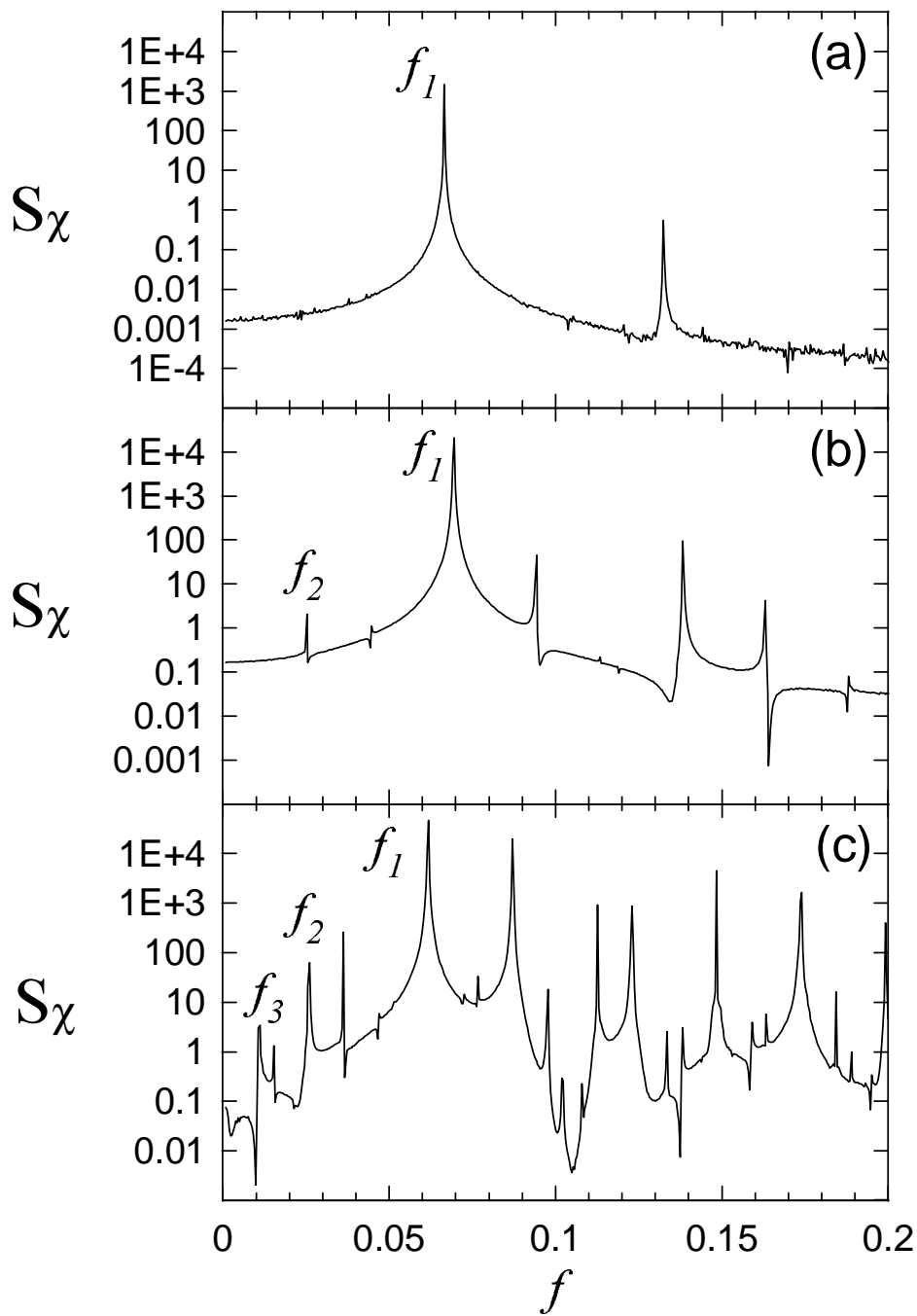
Fig. 8. Chirality spectrum for the time series corresponding to $E=3$, $%
\Delta _{1}=1.86$ (a), $\Delta _{1}=1.93$ (b) and $\Delta _{1}=1.99$ (c);
that correspond to regions $P$, $T$, and $T^{2}$ in the diagram in Fig.6.

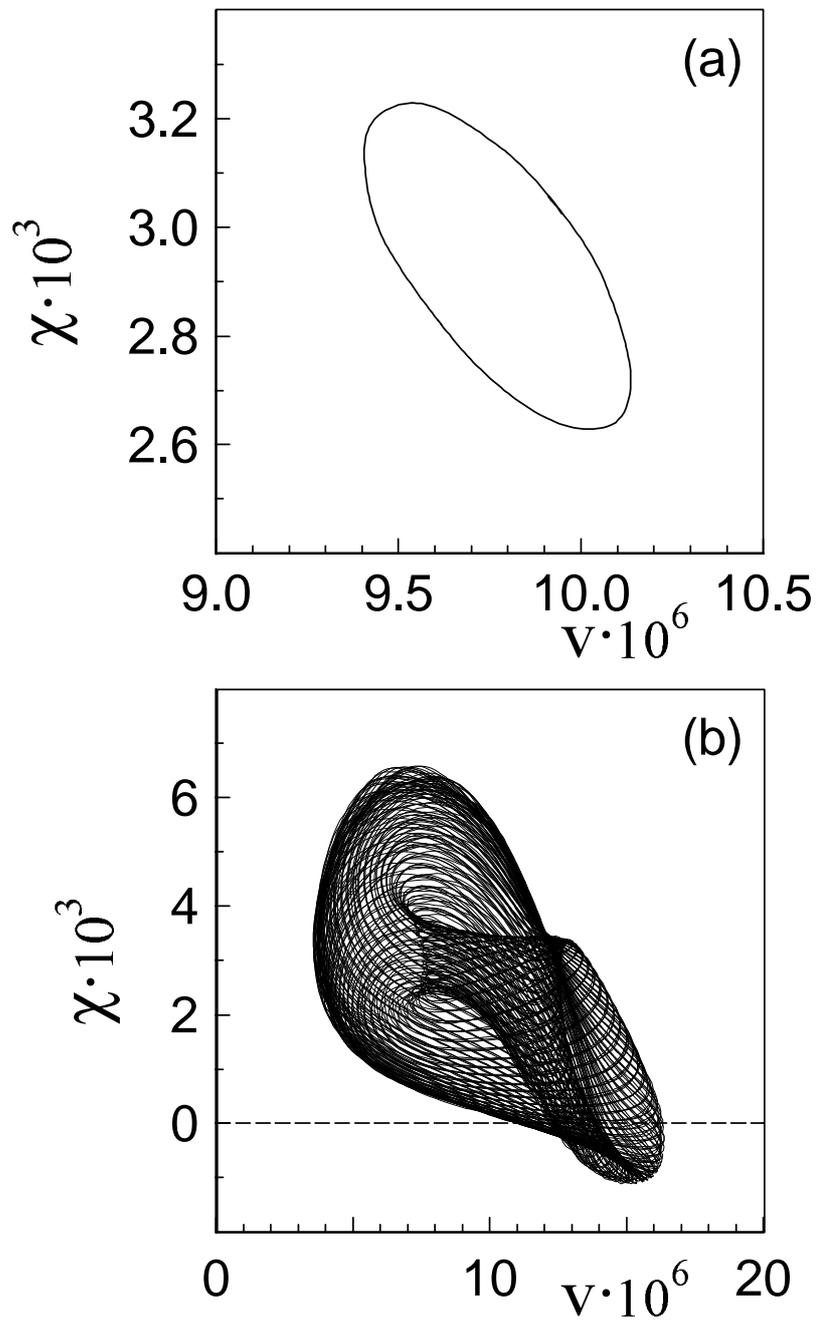
Fig. 9. Attractor projection on the $\left\langle v,\chi \right\rangle $
plane for (a) $E=3$ and $\Delta _{1}=1.86$ (periodic attractor) and $E=3$
and $\Delta _{1}=1.99$ (quasiperiodic attractor).

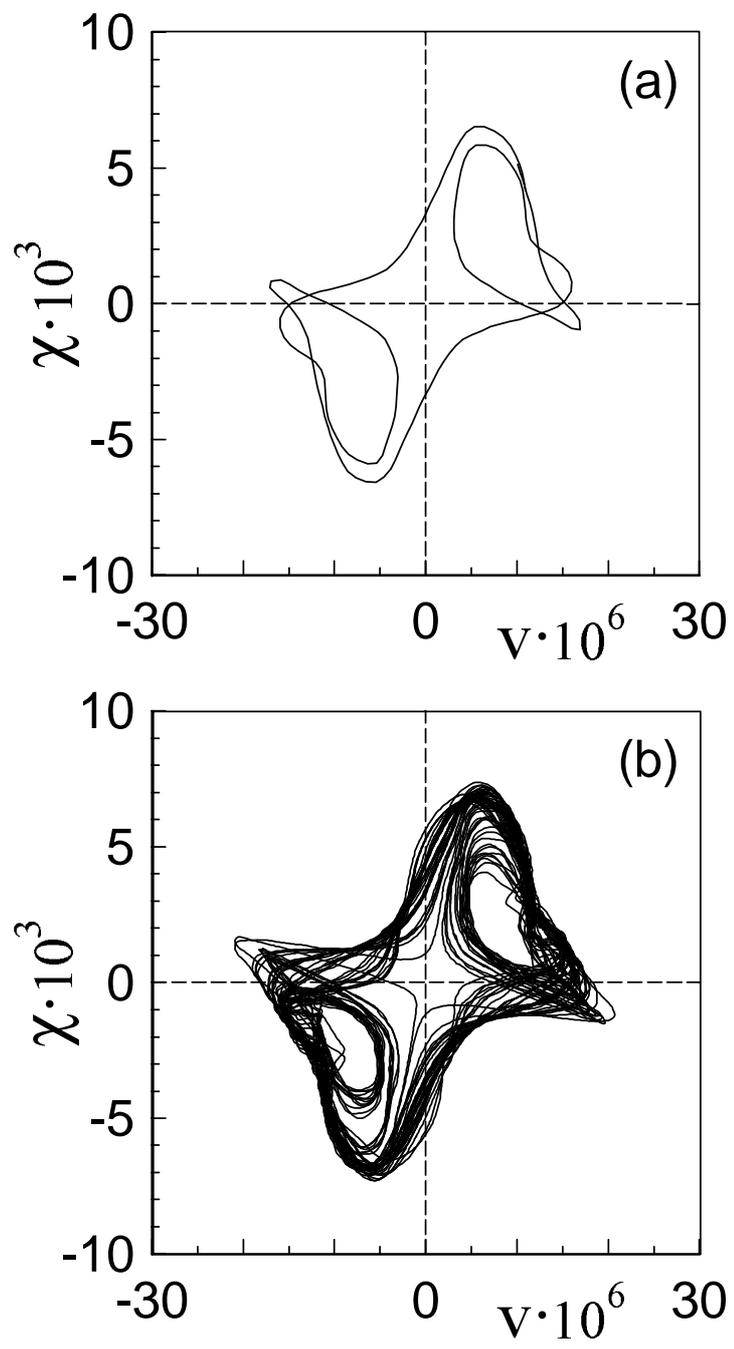
Fig. 10. Attractor projection on the $\left\langle v,\chi \right\rangle $
plane for (a) $E=3$ and $\Delta _{1}=2.02$ (periodic two--sided attractor)\
and (b) $E=3$ and $\Delta _{1}=2.04$ (chaotic attractor) corresponding to
regions $P$ and $C$, in the inset in Fig.6.

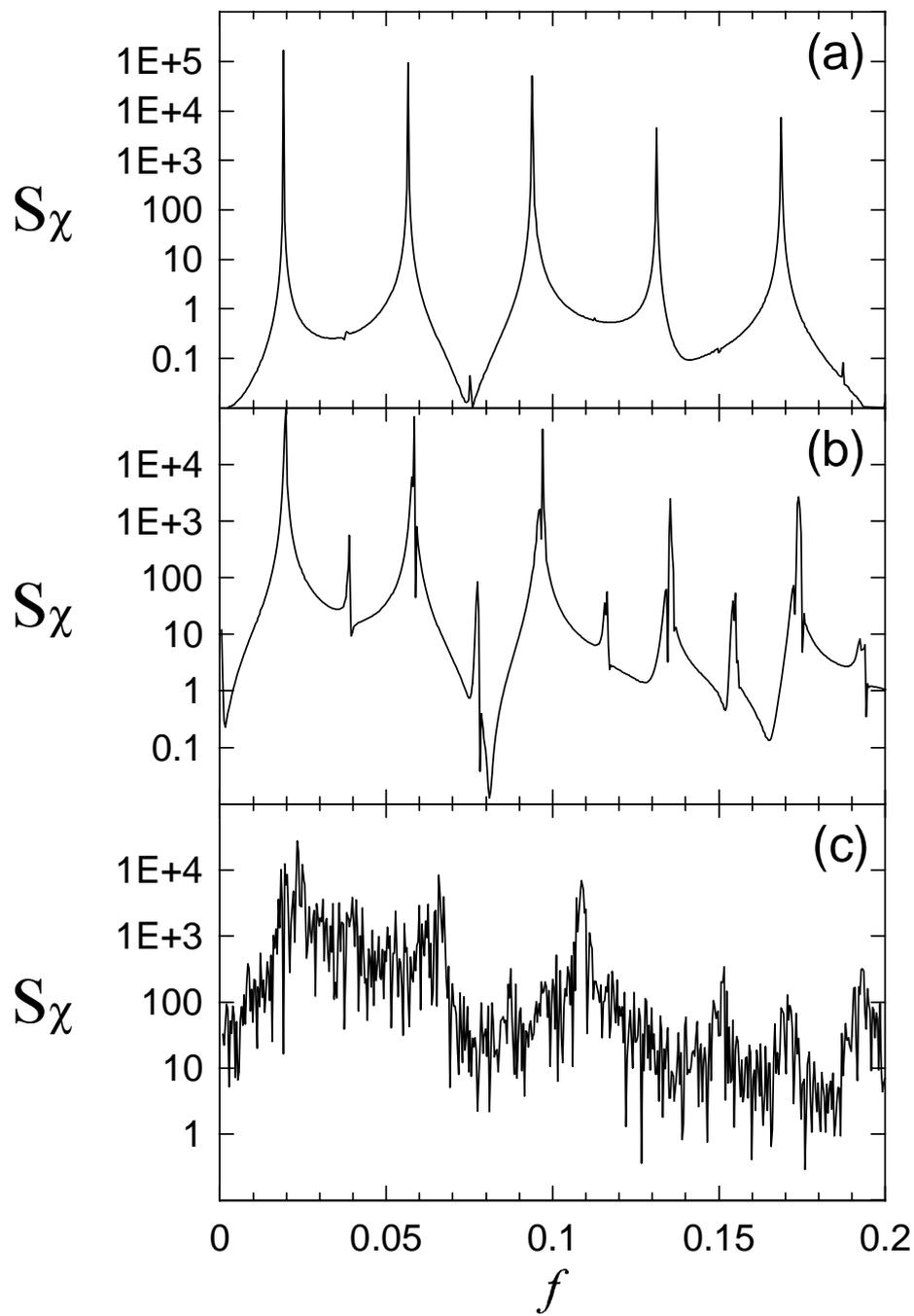
Fig. 11. Chirality spectrum for the time series corresponding to $E=3$ and $%
\Delta _{1}=2.02$ (a), $\Delta _{1}=2.0251$ (b) and $\Delta _{1}=2.04$ (c);
corresponding to regions $P$, $T$, and $C$ in the inset diagram in Fig. 6.

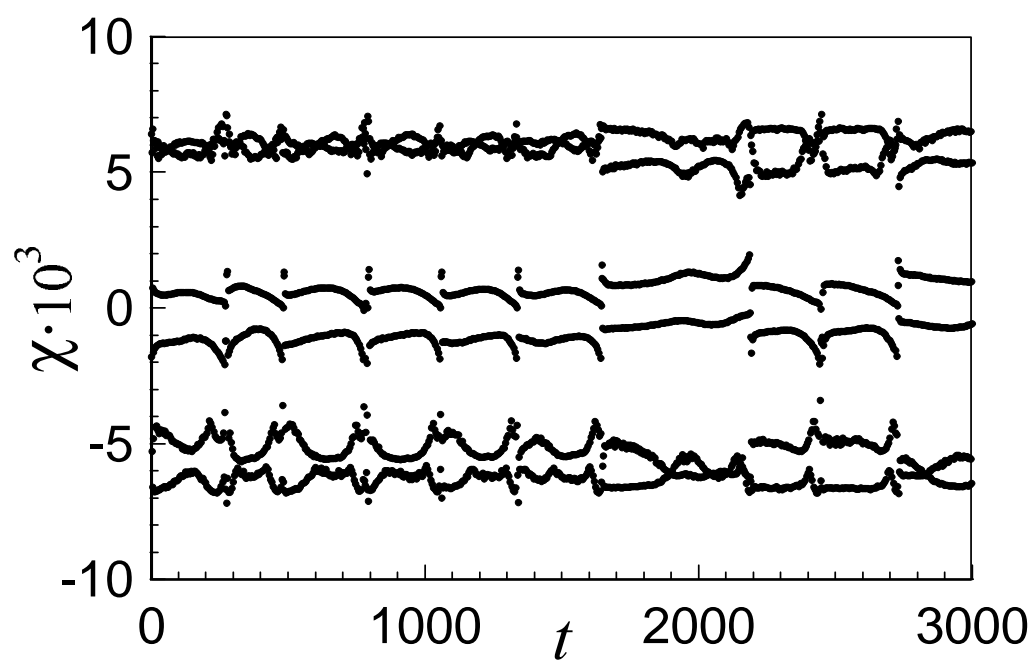
Fig. 12. Series of chirality extrema for $E=3$ and $\Delta _{1}=2.0260$,
corresponding to the intermittent route to chaos in the inset in Fig.6.

\end{document}